\documentstyle[editedvolume]{crckapb} 
\input{psfig}
\def\beq#1{\begin{equation}#1\end{equation}}
\def\bea#1{\begin{eqnarray}#1\end{eqnarray}}
\def\be{\begin{equation}}
\def\ee{\end{equation}}
\def\ba{\begin{eqnarray}}
\def\ea{\end{eqnarray}}
\def\d{{\rm d}}
\def\vx{{\bf x}}
\def\vk{{\bf k}}
\def\vg{{\bf g}}

\def\rhob{\overline{\rho}}
\def\deltam{\delta_{\rm mass}}
\def\deltal{\delta_{\rm lin.}}
\def\mg{\big<}
\def\md{\big>}
\def\mA{{\cal A}}
\def\mD{{\cal D}}
\def\mS{{\cal S}}

\def\mI{{\cal I}}

\def\mR{{\cal R}}
\def\mU{{\cal U}}
\def\Id{{\rm Id}}

\def\valpha{\vec{\alpha}}
\def\vbeta{\vec{\beta}}
\def\disp{\displaystyle}
\def\grad{\nabla}
\def\ii{\rm i}

\def \aa    #1 #2   {{\em Astr. \& Astrophys. \/} {\bf #1}, {#2}}

\def \aas   #1 #2   {{\em Astr. Astrophys. Suppl. Ser. \/} {\bf #1}, {#2}}
\def \aj    #1 #2   {{\em Astron. J. \/} {\bf #1}, {#2}}
\def \apj   #1 #2   {{\em Astrophys. J. \/} {\bf #1}, {#2}}

\def \apjs  #1 #2   {{\em Astrophys. J. Suppl. Ser. \/} {\bf #1}, {#2}}
\def \araa  #1 #2   {{\em Annual Review of Astr. \& Astrophys. \/} {\bf #1}, {#2}}
\def \mnras #1 #2   {{\em Mon. Not. R. astr. Soc. \/} {\bf #1}, {#2}}

\def \nat   #1 #2   {{\em Nature \/} {\bf #1}, {#2}}
\def \prevd   #1 #2   {{\em Phys. Rev. D \/} {\bf #1}, {#2}}

\begin{opening}
\title{Gravitational Lenses}
\author{Francis Bernardeau}
\institute{Service de Physique Th\'eorique, C.E. de Saclay\\
            F-91191 Gif-sur-Yvette Cedex, France}
\end{opening}
\runningtitle{Gravitational Lenses}

\begin{document}
\section{Introduction}

Gravitational lenses  are  becoming a precious mean   of probing  the matter
distribution in the Universe. From the search of matter in the Galactic halo
to the   study   of the    large-scale structures  of   the Universe,    the
gravitational  lens effects offer a unique  alternative to light surveys and
are now widely used.  This evolution is due in particular  to the use of new
observation devices, such as the wide field CCD cameras.

The aim of  this course is  to study the  effects of gravitational lenses in
those different astrophysical contexts.  These notes are voluntarily focused
on  the fundamental mechanisms  and  the basic  concepts that  are useful to
describe these effects. The  observational consequences will be presented in
more details in these  proceedings by Y. Mellier  (see also his review paper
1998). Related textbooks are,
\begin{itemize}
\item ``Gravitation'' by Misner, Thorne and Wheeler (1973) 
for General Relativity and in particular for the presentation of the
geometric optics.
\item ``Large-scale Structures of the Universe'' by Peebles (1993)
for the description of the large-scale structures of the Universe. 
\item ``Gravitational lenses'' by Schneider, Ehlers \& Falco (1992)
for a general (but rather mathematical) exhaustive presentation
of the lens physics.
\end{itemize}

 The content  of  these notes is   the  following.  In the   first section I
describe of  the  basic mechanisms of   gravitational lenses, techniques and
approximations that are usually employed.  The second  section is devoted to
the case of a  very simple deflector,  a point-like mass distribution.  This
corresponds  to   microlensing events in which  the   deflectors are compact
objects of  a fraction of a solar  mass  that may populate   the halo of our
Galaxy.  The last  two  sections are  devoted to cosmological  applications.
After  a  presentation of the  geometrical  quantities that  are specific to
cosmology, I will present the various phenomena that can be observed in this
context.  Finally I  describe the  weak  lensing regime.  This  is a rapidly
developing area that should eventually allow us to map the mass distribution
in  the   Universe.  I explore    how  this can be   used   to constrain the
cosmological parameters.

\section{Physical mechanisms}

\begin{figure}
\vspace{9.5 cm}
\special{hscale=70 vscale=70 voffset=0 hoffset=50 psfile=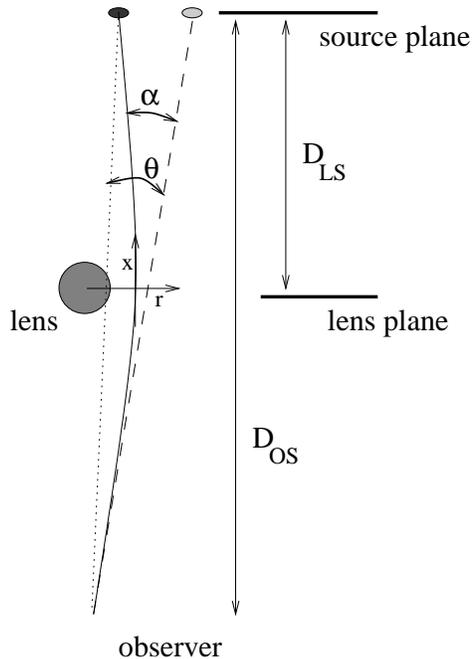}
\caption{The geometrical relationship between the deflection angle $\theta$
and the displacement angle $\beta$.
}
\label{deflec}
\end{figure}

The physical mechanisms  of gravitational lenses are   well known since  the
foundation   of General Relativity.   Any   mass concentration  is going  to
deflect photons that are passing  by with a fraction  angle per unit length,
$\delta \theta/\delta   s$, given  by 
\beq{ {\delta  \vec{\theta}\over\delta
s}=-2\,\vec{\nabla}_x{\phi\over c^2} 
} 
where the spatial derivative is taken
in a  plane that is  orthogonal to the photon  trajectory  and $\phi$ is the
Newtonian potential\footnote{We will see in section  \ref{cosmo} what is its
meaning in a cosmological context.}.

\subsection{Born approximation and thin lens approximation}

In practice,  the total deflection angle is at most about an
arcmin. This is the case for the most massive   galaxy clusters. It
implies that in the subsequent calculations it is possible to ignore
the bending of the trajectories and calculate the lens effects as if
the trajectories  were straight lines. This is the Born approximation. 

Eventually, one can do another approximation by noting that in general
the deflection takes place along a very small fraction of the
trajectory between the sources and the observer. One can then assume
that the lens effect is instantaneous and is produced through the
crossing of a plane, the lens plane. This is the thin lens
approximation.

\subsection{The induced displacement}

The direct consequence of this bending is a displacement of the
apparent position  of the background objects. This apparent
displacement depends on the  distance of the source plane, $D_{OS}$,
and on the distance between the lens plane and the source plane
$D_{LS}$.  More precisely we have (see Fig. \ref{deflec}), 
\beq{
\vec{\beta}=\vec{\alpha}-{2\over c^2}\,{D_{LS}\over D_{OS}\,D_{OL}}
\,\vec{\nabla}_{\alpha}\left(\int \d s\ \phi(s,\alpha)\right) 
} where
$\vec{\alpha}$ is the position in the image plane,  $\vec{\beta}^S$ is
the position in the source plane. The gradient is taken here with
respect to the angular position (this is why a $D_{OL}$ factor
appears). The total deflection is obtained by an integration along the
line of sight, assuming the lens is thin.  In  a cosmological context
the exact expressions  of the angular distances are not trivial, they
depend on the local curvature of the background. 

\section{The case of a point-like mass distribution}

\subsection{Multiple images and displacement field}

\begin{figure}
\centerline{
\psfig{figure=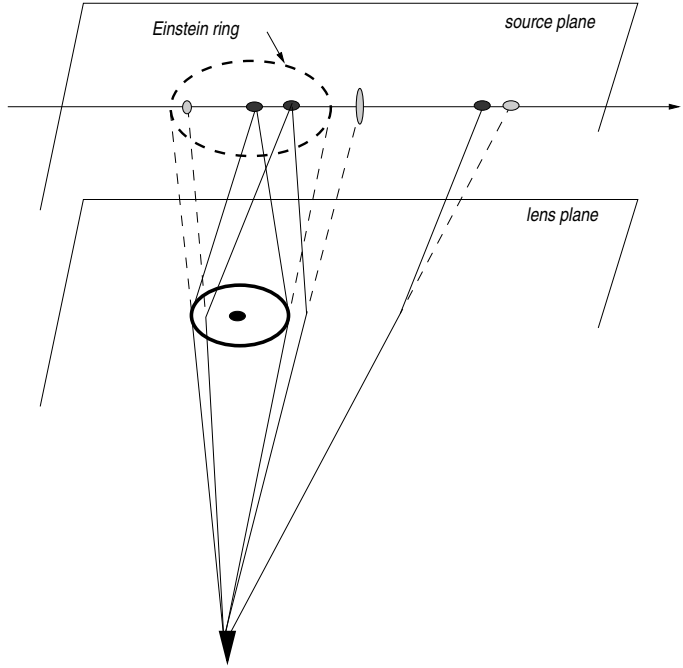,width=9cm}}
\caption{Position of the Einstein ring for a point-like mass distribution.
}
\label{crit}
\end{figure}

The potential of a point-like mass distribution is given by, 
\beq{
\phi(r)={-G\,M\over r}, 
} for an object of mass $M$.  Let me calculate
the instantaneous deflection angle at an apparent distance $r$. We
suppose that the impact  parameter of the trajectory is $r$ and $\vx$
is the abscissa to the point of the trajectory  that is the closest to
the lens (see  Fig. 1).  Along the trajectory the potential is given
by, 
\beq{ \phi(x)={-G\,M\over \sqrt{r^2+x^2}}. 
} Then the deflecting
angle is given by, 
\be {\delta \theta\over\delta x}=-2\,{G\,M \,r\over
c^2\,(r^2+x^2)^{3/2}}.  
\ee 
The total deflection angle $\theta$ is
given by the result of the integration of this quantity with respect
to $x$. It gives, 
\beq{ \theta={4\,G\,M\over r\,c^2}.  
} This is a
well known expression which served as the first test of General
Relativity  with the observed displacement of stars around the solar
disc (Dyson et al. 1919).  

It implies that the true position of an object on the sky, $\beta$, is 
related to its apparent  position, $\alpha$, with
\beq{
\vbeta=\valpha-{R_E^2\over \alpha^2}\,\valpha
}
where $\vbeta$ and $\valpha$ are 2D angular position vectors and $R_E$
is the Einstein radius,
\beq{
R_E=\sqrt{{4\,G\,M\over c^2}\,{D_{LS}\over D_{OS}\,D_{OL}}}.
}
One can see that when $\alpha^2=R_E^2$ the lens and the background objects
are necessarily aligned. It implies that, since the optical bench is
symmetric 
around its axis, the observed object appears as a perfect ring (see
Fig. \ref{crit}). It is worth noting that for this potential, 
except for this particular position, all background objects have
two images. This is however quite specific to
a point like mass distribution which has a singular gravitational potential.

The problem is that none of these features are observable when the
lens is a star.  Let for example assume that we have a one solar mass
star in the halo of our galaxy (therefore at a distance of about 30
$kpc$).  The apparent size of such a star is about $10^{-8}$
arcsec. Its Einstein ring is about $10^{-4}$ arcsec\footnote{To do
this calculation it is  useful to know that the horizon of a one solar
mass black hole, $r=2\,G\,M_{\odot}/c^2$, is about 3 km.}.  None of
these dimensions are accessible to the observations (the angular
resolution of telescope is at best a few tens of arcsec).  The
Einstein radius is therefore much too small to be actually seen!

Note however that these numbers show that the point-like approximation
is entirely justified for a star (the Einstein ring is much more bigger that 
the apparent size of a star). Simple examination of the scaling
in those relations shows that this would not be true for massive
astrophysical objects such as galaxies or galaxy clusters.

The detection of gravitational effects  due to stars should then be
done by another mean: the amplification effect.

\subsection{The amplification matrix}

The case of circular lenses has already given us a clue: when the
source is precisely aligned with the lens, the image is no more a
point but a circle. One consequence is that the observed total
luminosity is much larger than what would have been observed without
lenses. The effect is basically due to the variations of the
displacement  field with respect to the apparent position. These
variations induce a change  of both the size and shape of the
background objects. To quantify this effect one can compute  the
amplification matrix $\mA$ which describes the linear change between
the source plane and the image plane, 
\beq{ \mA=\left({\partial
\alpha_i\over \partial \beta_j}\right).  
} Its inverse, $\mA^{-1}$, is
actually directly calculable in terms of the gravitational
potential. It is given by the derivatives of the displacement with
respect to the apparent position, 
\beq{ \mA^{-1}\equiv {\partial
\beta_i\over \partial \alpha_j} =\delta_{ij}-2{D_{LS}\over
D_{OS}\,D_{OL}} \phi_{,ij}.
\label{AmpEuc}
}

In case of a point-like mass distribution it is easy to see that,
\beq{
\mA^{-1}=\left(\delta_{ij}\left[1-{R_E^2\over
\alpha^2}\right]+{\alpha_i\alpha_j\over \alpha^2} 
R_E^2\right).
}
The amplification effect for each image is given by the inverse of the
determinant 
of the amplification matrix computed at the apparent position of the
image. The amplification factor is usually noted $\mu$,
\beq{
\mu=1/\det(\mA^{-1}).
}
In case of the point-like distribution we have
(the calculation is simple at the position $\alpha_1=\alpha,\ \alpha_2=0$),
\beq{
\mu=\left\vert{\alpha^4\over \alpha^4-R_E^4}\right\vert,
}
{\it for each image}. The {\it total} amplification effect is given by the
summation of the two effects for the 2 images, 
\beq{
\mu_{\rm tot}={u^2+2\over u(u^2+4)^{1/2}}\ \ {\rm with}\ \ u={b\over R_E},
}
where $b$ is the impact parameter of the background object in the
source plane. The amplification effect is obviously dependent on the 
impact parameter. If it is changing with time, this effect is detectable.

\subsection{The microlensing experiments}

The microlensing experiments are based on this effect. When a compact
object of the halo of our galaxy reaches, because of its proper
motion,  the vicinity of the light path of a background star (from the
SMC or the LMC) the impact parameter is changing with time and can be
small enough to induce a detectable amplification (when $u$ is about
unity, the amplification is about 30\%).  In practice one observes
changes in the magnitude of the remote stars that obey specific
properties,
\begin{itemize}
\item the time dependence of the amplification is symmetric and has a
specific shape; 
\item the amplification effect is unique;
\item the magnitude of the amplification effect is the same in all
wavelengths. 
\end{itemize}

The time scale of such an event is about a few days to a few month
depending on the mass of the deflectors. Currently a fair number of
such events have been recorded (see contribution of J. Rich, these
proceedings) and constraints on the content    of our halo with low
massive compact objects have been put.

\section{Gravitational lenses in Cosmology}
\label{cosmo}

The extension of the lens equations to a cosmological context
raises some technical difficulties because the background
in which the objects are embedded is not flat. 
The aim of this section is to clarify these points. However, readers 
that are not familiar with cosmology can jump to section \ref{clusters}.

The basic equations, that describe jointly the evolution
of the expansion parameter and the mean density, are the following,
\bea{
3\, \ddot{a}&=&-4 \pi G\,\rho\,a+\Lambda\,a\label{Evola1};\\
\dot{a}^2&=&{1\over a}
\left[{8\pi\,G\over 3}\,\rho\,a^3-k\,a
+{\Lambda\,a^3\over 3}\right]\label{Evola2}.
}
where $\Lambda$ is a possible cosmological constant
and  $k$ a possible curvature term.
The Hubble constant reads,
\beq{
H={\dot{a}\over a}=\sqrt{{8\pi\,G\over 3}\,\rho-{k\over a^2}+
{\Lambda\,a^3\over 3}}.
}
To simplify the discussions the reduced quantities are introduced,
\beq{
\Omega={8\pi\,G\over 3\,H^2}\,\rho\ \ {\rm and}\ \ 
\lambda={\Lambda\over 3\,H^2}.
\label{OmegaDef}
}
They have an index $0$ when they are taken at present time.

\subsection{The angular distances}

We consider an object of size $l$ (either because of its
proper size or because of a peculiar physical process) at redshift $z$.
When this size $l$ is seen under an angle $\alpha$, then by definition,
\beq{
\alpha={l\over \mD_0},
}
where $\mD_0$ is the angular distance. This is the distance at which this
object would be in an Euclidean metric. What is then the relationship
between  $\mD_0$ and $z$? The Friedmann-Robertson-Walker metric is given by,
\beq{
\d s^2=-c^2\,\d t^2+a^2(t)\left(
{\d x^2\over1-k\,x^2}+x^2\d\theta^2+x^2\,\sin^2\theta\,\d\varphi^2\right).
}
The fact that the size of this object is $l$ means that
it takes a time interval $\d t=l/c$ for light to travel 
to one end to the other.
The corresponding angle $\alpha=\d \theta$ can be obtained
by writing $\d s=0$ with $\d x=0, \d \varphi=0$, which gives
\beq{
l=a(t)\,x\,\alpha.
}
The angular {\em comoving} distance is thus given by
$\mD_0(z)=x(z)$.
The expression of $x(z)$ can then be computed by the relation
$\d s=0$ along  the line of sight with  $\d\theta=0$ and $\d\varphi=0$,
\beq{
\int_{t_0}^{t_1}{c\,\d t\over a}=
\int_0^{x(z)}{\d x\over \sqrt{1-k x^2}}\equiv \chi(z).
\label{distang}
}
For an open Universe, $k<0$, and we have,
\beq{x(z)=\mD_0(z)={1\over\sqrt{-k}}\sinh[\sqrt{-k}\ \chi(z)].
}
Obviously when $k=0$, $x=\chi$.

The relation $\chi(z)$ depends on the function $a(t)$ for a given
cosmology, and therefore on the matter content of the Universe, on 
$\Lambda$ and $k$ .
For instance, for an Einstein-de Sitter Universe, in the matter
dominated era, we have
\beq{
a(t)=\left({t\over t_0}\right)^{2/3},
}
so that,
\beq{
x(z)=\int_{t_0}^{t_1}t^{-2/3}\d t=3(t_0^{1/3}-t_1^{1/3})t_0^{2/3},
\ \ H_0={2\over 3\, t_0},
}
and eventually,
\beq{
\mD_0(z)=\chi(z)={c\over H_0}\left(2-{2\over\sqrt{1+z}}\right).
}
This is the comoving angular distance for an Einstein-de Sitter
Universe. More generally we have an explicit solution if $\Lambda=0$ only.
Finally the lens equation also requires the
angular distance between two different redshifts $z_1$ and $z_2$.
The calculation is actually quite simple. This distance is given
formally by $\mD_0(z_2)$ when it is calculated at a time when
the observer is at redshift $z_1$. $k$ being time independent we
should formally have,
\beq{
\mD_0(z_1,z_2)={1\over \sqrt{-k}}\sinh[\sqrt{-k}\ (\chi(z_2,z_1))],
}
and to compute  $\chi(z_2,z_1)$ one only needs to remark,
\beq{
\chi(z_2,z_1)\equiv\int_{t_1}^{t_2}{c\,\d t\over a}=\chi(z_2)-\chi(z_1),
}
which gives the expression of the angular distance we need.
Eventually, it is fruitful to notice that,
\beq{
{c\,\d t\over a}=-{c\,\d z\over H},
}
which gives,
\beq{
\mD_0(z_1,z_2)={c\over H_0\sqrt{1-\Omega_0-\lambda_0}}\sinh\left[
H_0\sqrt{1-\Omega_0-\lambda_0}\ \int_{z_1}^{z_2}{\d z\over H(z)}\right],
}
and
\beq{
\mD_0(z)={c\over H_0\sqrt{1-\Omega_0-\lambda_0}}\sinh\left[
H_0\sqrt{1-\Omega_0-\lambda_0}\ \int_{0}^{z}{\d z'\over H(z')}\right].
}
The whole geometrical part of the lens equation
is thus established.

\subsection{Geometric optics in a weakly inhomogeneous Universe}

What is now the source term for the deflection angle?
We should first notice that in absence of lenses the light rays 
follow the geodesics of the Friedmann-Robertson-Walker metric. 
And in the applications we are interested in, 
the metric fluctuations are always weak. These fluctuations
are given by, ${G\,M/ (R\,c^2)}$. For instance, for
\begin{itemize}
\item
1 star: $M=1\,M_{\odot},\ R=7\,10^5{\rm km},\ \delta\phi\approx 10^{-6}$;
\item
1 galaxy cluster: $M=10^{15}\,M_{\odot},\ R=1{\rm Mpc}=3\,10^{19}{\rm km},\ \delta\phi\approx 10^{-5}$.
\end{itemize}
The metric inhomogeneities are thus always extremely weak, even in the
most extreme cosmological situations.

\begin{figure}
\centerline{
\psfig{figure=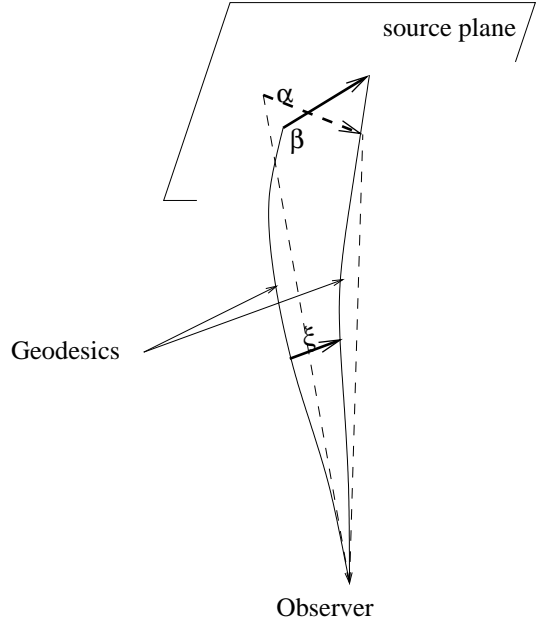,width=7cm}}
\caption{Sketch showing the geometrical quantities that appear in 
Eq. \ref{dij}
}
\label{Sachs}
\end{figure}

Following\footnote{See Misner, Thorne and Wheeler for an exhaustive
presentation of the geometric optics.} Sachs (1961), 
we consider two nearby geodesics, ${\cal L}$ and
${\cal L'}$, in a light bundle in an FRW Universe with small metric
fluctuations. We denote
$\alpha_i$ the bi-dimensional angular distance between ${\cal L}$
and ${\cal L'}$ as it is seen by the observer. This is the distance
in the image plane, that is the difference between the angular
coordinates with which the photons arrive. We denote 
$\xi_i(z)$ the real distance between ${\cal L}$ and ${\cal L'}$
at redshift $z$ (see Fig. \ref{Sachs}). It implies that 
the geodesics are straight enough so that light
always travels towards the observer. We also assume that the deflections
are small enough so that it is possible to make
the small angle approximation,
\beq{
\xi_i(z)=a\ \mD_{ij}(z)\ \alpha_j,
\label{dij}
}
that is that we assume the the position vector $\xi_i$ can be
obtained by a simple linear transform of the angular
coordinates. For an homogeneous space $\mD_{ij}(z)$ is simply
given by $\mD_0(z)\ \delta^{K}_{ij}$ where $\delta^{K}_{ij}$ is
the Kronecker symbol. Obviously $\mD_{ij}$ changes
as a function of redshift along the trajectories. The ``virtual''
angular position in the source plane is then given
by the ratio of the real distance (at time of light
emission for instance) by the angular distance of the emitter
in an homogeneous space,
\beq{
\vec\beta={\vec\xi(z)\over a \mD_0(z)}.
}
The amplification matrix, or rather its inverse,  $\mA^{-1}$, is
then given by,
\beq{
\mA^{-1}(z)={\mD_{ij}(z)\over a \mD_0(z)}.
}
for a source plane at  redshift $z$.

Sachs (1961) gave the master equation which governs the
evolution of the distance between the geodesics. The derivation of this
equation goes beyond these lecture notes and I give only
the final answer (as it has been given by Seitz, Schneider and Ehlers 1994),
\beq{
{\d^2[a\ \mD_{ij}(\vec\beta,z)]\over \d\eta^2}=a(z)\mR_{ik}(\vec\beta,z)\ 
\mD_{kj}(\vec\beta,z)
\label{sachs}
}
where the derivatives are taken with respect to $\eta$, 
\beq{
\d\eta=-{\d a\over H(a)}=-a\ \d t,\ \ \ \eta(z=0)=0.
}
with the boundary conditions,
\beq{
\left(\mD_{ij}\right)_{z=0}=0;\ \  
\left({\d \mD_{ij}\over \d\eta}\right)_{z=0}={c\over H_0}.
}
The matrix  $\mR_{ij}$ represents the tidal effects. It can by
written in terms of the gravitational potential $\phi$ given by,
\beq{
\Delta_{x}\phi=4\pi\,G\,\overline{\rho}\,a^2\,\deltam.
\label{PoissonCosmo}
}
The Laplacian is taken with respect of the comoving angular
distances. We have
\beq{
\mR_{ij}=-{4\pi\,G\,\overline{\rho}\over H_0^2\,a^{2}}\left(
\begin{array}{cc}
1&0\\
0&1
\end{array}
\right)-{2\over H_0^2\,a^{2}}\left(
\begin{array}{cc}
\phi_{,11}&\phi_{,12}\\
\phi_{,21}&\phi_{,22}
\end{array}
\right).
}
Since $8\pi\,G\,\overline{\rho}\,a^3=3H_0^2\,\Omega_0$, (e.g. Eq.
\ref{OmegaDef}) for an homogeneous Universe we have,
\beq{
\mR_{ij}^{(0)}=-{3\over 2}\,(1+z)^5\,\Omega_0\,\delta^K_{ij}
}
(the superscript (0) means here that it is the value
of $\mR$ without perturbations). In this case the matrix
$\mD_{ij}$ is proportional to $\delta^K_{ij}$ and we have,
\beq{
{\d^2[a\ \mD_0(z)]\over \d\eta^2}=-{3\over 2}
(1+z)^4\,\Omega_0\,\mD_0(z).
}
We recover in fact the comoving angular distance the expression of which we
know,
\beq{
\mD_0(z)={c\over H_0\sqrt{1-\Omega_0-\lambda_0}}\,
\sinh\left[\sqrt{1-\Omega_0-\lambda_0}\int_0^z{\d z'\over E(z')}\right],
}
with
\beq{
E(z)={H(z)\over H_0}=
\sqrt{\lambda_0+(1+z)^2(1-\Omega_0-\lambda_0)+(1+z)^3\Omega_0}.
}
This integral has a closed form only when the cosmological
constant, $\Lambda$, is zero.

\subsection{The linearized equation of geometric optics}

We can remark that the equation (\ref{sachs}) is not linear since
$\mD$ is not simply proportional to  $\mR$. This expresses the fact that
the deformation of the angular distance is made all along the
light trajectory by multiple deflections. The general resolution of
Eq. (\ref{sachs}) 
is in general very complicated. It can however
handled when it is linearized. Let's assume we can
expand  $\mD_{ij}$ with respect of the local density contrast,
\beq{
\mD_{ij}(z)=\mD_{0}+\mD_{ij}^{(1)}+\dots
}
It implies that, at first order,
\beq{
{\d^2[a\ \mD^{(1)}_{ij}(\vec\beta,z)]\over \d\eta^2}-
a(z)\mR_{ik}^{(0)}(\vec\beta,z)\mD_{kj}^{(1)}(\vec\beta,z)
=-{3}\Omega_0(1+z)^4\mD_0(z)\varphi_{,ij}(\vec\beta,z)
}
with,
\beq{
\left(\mD_{ij}^{(1)}\right)_{z=0}=0,
\ \ 
\left({\d\mD_{ij}^{(1)}\over\d \eta}\right)_{z=0}=0,
}
and we define the field $\varphi$ so that,
\beq{
\Delta_x\varphi=\delta_{\rm mass}(\vec\beta,z)=
{\Delta_x\phi\over4\pi\,G\,\overline{\rho}\,a^2}.
}

To solve this differential equation it is easier to write
it with the variable $z$. It then reads,
\bea{
{\d^2[a\ \mD^{(1)}_{ij}(\vec\beta,z)]\over \d z^2}+{1\over E(z)}{\d E(z)
\over \d z}{\d \mD^{(1)}_{ij}(\vec\beta,z)\over \d z}-\nonumber\\
-{1\over 1+z}{1\over E(z)}{\d E(z)
\over \d z}\mD^{(1)}_{ij}(\vec\beta,z)+{3\over 2}{\Omega_0 (1+z)\over E^2(z)}
\mD^{(1)}_{ij}(\vec\beta,z)=\nonumber\\
-{3}
{\Omega_0 (1+z)\over E^2(z)} \mD_0(\vec\beta,z)\varphi_{,ij}(\vec\beta,z).
\label{SachsLin}
}
The differential homogeneous equation it is associated with
has two known solutions. One describes the angular distance
$\mD_0$, the other is given by,
\beq{
\mU_0(z)=
{1\over \sqrt{1-\Omega_0-\lambda_0}}\,
\cosh\left[\sqrt{1-\Omega_0-\lambda_0}\int_0^z{\d z'\over E(z')}\right],
}
The general solution of \ref{SachsLin} then reads,
\beq{
\mD_{ij}^{(1)}(\vec\beta,z)=-3\Omega_0
\int_0^z\d z'{(1+z')\mD_0(z')\varphi_{,ij}(z')\over
E^2(z)} {\mU_0(z)\mD_0(z')-\mU_0(z')\mD_0(z)\over
\mU_0'(z')\mD_0(z')-\mU_0(z')\mD_0'(z')},
}
which, after elementary mathematical transforms, gives,
\bea{
\mD_{ij}^{(1)}(\vec\beta,z)=-3\,{\Omega_0}
\int_0^z{\d z'\over E(z)}{1\over \sqrt{1-\Omega_0-\lambda_0}}\times\nonumber\\
\times
\sinh\left[\sqrt{1-\Omega_0-\lambda_0}\int_z^{z'}{\d z''\over E(z'')}\right]
(1+z')\mD_0(z')\varphi_{,ij}(z').
}
It can be rewritten by introducing the physical distance $\chi$ along
the line of sight. We eventually have,
\bea{
\mA^{-1}(z)=\Id-{3\,\Omega_0\over (c/H_0)^2}
\int_0^{\chi(z)}\d \chi' {\mD_0(z',z)\mD_0(z')\over\mD_0(z)}
(1+z')\varphi_{,ij}(z'),
\label{AmpSachs}
}
where the angular distances  $\mD_0(z)$ and $\mD_0(z,z')$ are comoving.
This equation actually gives the expression of the amplification
matrix for a non-trivial background. We find that the amplification
matrix is given by  the superposition of lens effects of
the different mass layers. We can remark that the
lens term is given by the gravitational potential, $\phi$, that is
by the potential the source term of which is given by the density
contrast.

Note finally that this equation is valid in two limit
cases, either for a single lens plane with an arbitrary
strength or the superposition of any number of weak lenses.
This equation naturally extents the previous result, (\ref{AmpEuc}),
obtained for a single lens in an Euclidean background.
The higher orders of Eq. (\ref{sachs}) give the intrinsic
lens coupling effects (i.e. their nonlinear parts). We will not
consider them here.

\section{Galaxy clusters as gravitational lenses}
\label{clusters}

The study of galaxy clusters has become a very active field
since the discovery of the first gravitational arc by 
Soucail et al. (1988) in Abell cluster A370.
Galaxy clusters give the most dramatic example of
gravitational lens effects in a cosmological context.
The difficulty is however to describe the shape of their
mass distribution.

\subsection{The isothermal profile}

For an isothermal profile 
we assume that the local density $\rho(r)$ behaves like,
\beq{
\rho(r) = \rho_0  \left({r\over r_0}\right)^{-2}.
}
With such a density profile the total mass is not finite. So this is
not a realistic description but it is a good starting point for the
central part of clusters.
It is actually more convenient to parameterize the depth of
a potential well with the velocity dispersion it induces. 
The velocity dispersion is due to the random velocity that particles
acquire when they reach a sort of thermal equilibrium. 
Such a dispersion is in principle measurable
with the observed galaxy velocities along the line of sight.
The velocity dispersion is related to the mass $M(<r)$ of the
potential well that is included within a radius $r$,
\beq{
\sigma^2(r)\sim {G\,M(<r)\over r}.
}
In case of a isothermal profile, the 
velocity dispersion is {\it independent} of the radius
and we have
\beq{
\sigma^2=2\pi\,G\,\rho_0\,r_0^2.
}
The integrated potential along the line of sight is given by,
\beq{
\varphi(r)=2\pi\,\sigma^2\,r.
}
As a consequence the amplitude of the displacement is independent
of the distance to the cluster center and 
\beq{
\vbeta=\valpha-{4\pi\over c^2}
\,{D_{LS}\over D_{OS}}\sigma^2\,{\valpha\over \alpha}.
}
The position of the Einstein ring is obviously given by,
\beq{
\alpha=R_E={4\pi\over c^2}\,\,{D_{LS}\over D_{OS}}\sigma^2,
}
which depends both on the velocity dispersion and on the angular distances.
The number of images depends in this case on the value of the impact parameter.
If it is too large (i.e. larger than $R_E$) then each background object has only
one image.
For a galaxy cluster of a typical velocity dispersion of
$500\ km/s$, and for a source plane situated at twice the distance
of the lens, the size of the Einstein ring is about $0.5$ arcmin.
It is interesting to note that the size of the Einstein ring is 
directly proportional to the square of velocity dispersion (in units
of $c^2$) and to the ratio $D_{LS}/D_{OS}$. 

The amplification matrix reads,
\beq{
\mA^{-1}=\left(
\begin{array}{cc}
1&0\\
0&1-{1\over x}
\end{array}
\right),
}
where we have,
\beq{
x={r\over R_E}.
}
As a result the amplification is given by,
\beq{
\mu={x\over 1-x}.
}
Once again the amplification becomes infinite when $x\to 1$ that is,
close to the critical line.

\subsection{The critical lines for a spherically
symmetric mass distribution}

\begin{figure}
\vspace{9 cm}
\special{hscale=55 vscale=55 voffset=0 hoffset=0 psfile=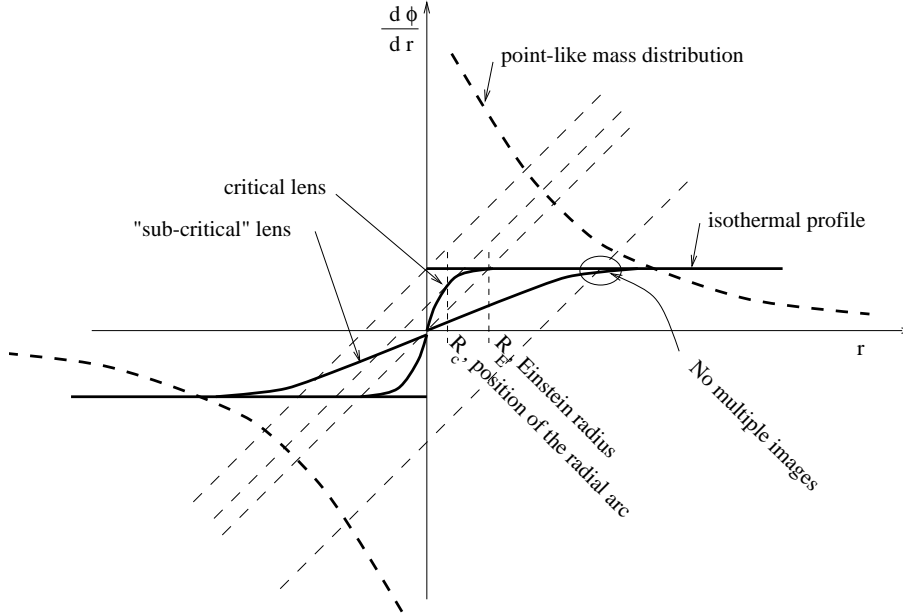}
\caption{Graphical determination of the position and number of images
from the shape of the potential.
}
\label{clines}
\end{figure}

The two previous cases correspond to specific profiles. In this part
I only assume a spherical symmetric profile.
The displacement is then given by the derivative of the potential,
\beq{
\vbeta=\valpha-{\d \varphi\over \d r} {\valpha\over \alpha}.
}
It is interesting to visualize this relation with
a graphic representation. This is proposed in Fig. \ref{clines}.
The number and position of the images of a given background object
are given by the number of intersection points between the curve and
a straight line of slope unity. This is a direct consequence of
the relation,
\beq{
b-a=-{\d \varphi(a)\over \d a}
}
when the potential is computed along a given axis
that crosses the cluster through the center and $a$ and $b$
are the abscissa on this axis of one given object in respectively the
source and the image plane.

The interesting quantity is also the amplification matrix that indicates
the position of the critical lines. In general this matrix reads, 
\beq{
\mA^{-1}=\left(\begin{array}{cc}
1-{\partial^2 \varphi\over \partial r^2} & 0\\
0 & 1-{1\over r}{\partial \varphi\over \partial r} 
\end{array}\right),
}
when it is written in the basis $(\vec{e}_r,\vec{e}_{\theta})$.
Then the amplification is infinite in two cases, when
\beq{
{\partial^2 \varphi\over \partial r^2}=1\ \ {\rm or}\ \ 
{1\over r}{\partial \varphi\over \partial r}=1.
}
The second eigenvalue corresponds to the same case as for a singular
isothermal profile. At this particular position the source forms
an Einstein ring. 
The first eigenvalue, however, is associated with an eigenvector that is along
the $x$ direction, that is along the radial direction. It means that 
the "arc" which is thus formed is radial. It graphically corresponds to the case
of two merging roots. It is therefore directly associated with the
behavior of the potential near the origin.

\subsection{The isothermal profile with a core radius}

Let us consider a simple case where the projected potential is made
regular near the origin,
\beq{
\varphi(r)=\varphi_0\,\sqrt{1+(r/r_c)^2}
}
The constant $\varphi_0$ is related to the velocity dispersion with
\beq{
\varphi_0={4\pi\sigma^2\over c^2}\,{D_{LS}\,D_{OL}\over D_{OS}\,r_c}
}
where $\sigma$ is here the velocity dispersion at a radius much larger that
$r_c$ (the velocity dispersion decreases to zero at the origin 
in this model). This is a more realistic case. It is interesting to
note that in this case the potential is not necessarily critical (there
may be no region of multiple images, see Fig. \ref{clines}). When 
it is critical the discovery of a radial arc is an extremely precious
indication for the value of the core radius.

\subsection{Critical lines and caustics in realistic mass distributions}

In realistic reconstructions of lens potential however, it is very
rare that the lens  is circular. Most of the time the mass
distribution of the lens is much more complicated. It induces complex
features and series of multiple images. 

\begin{figure}
\vspace{9 cm}
\special{hscale=40 vscale=40 voffset=150 hoffset=0 psfile=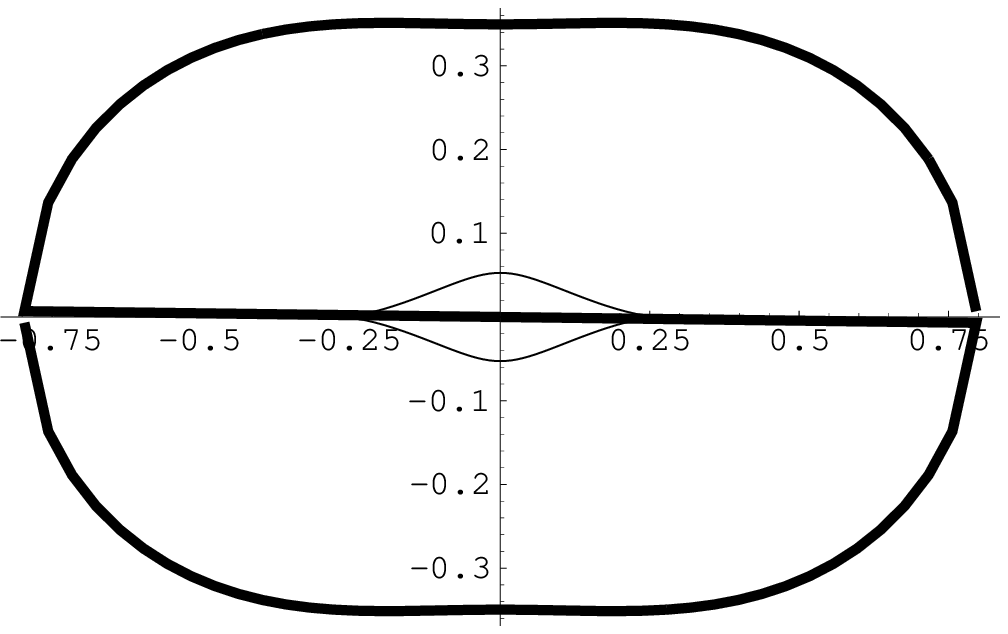}
\special{hscale=50 vscale=50 voffset=130 hoffset=150 psfile=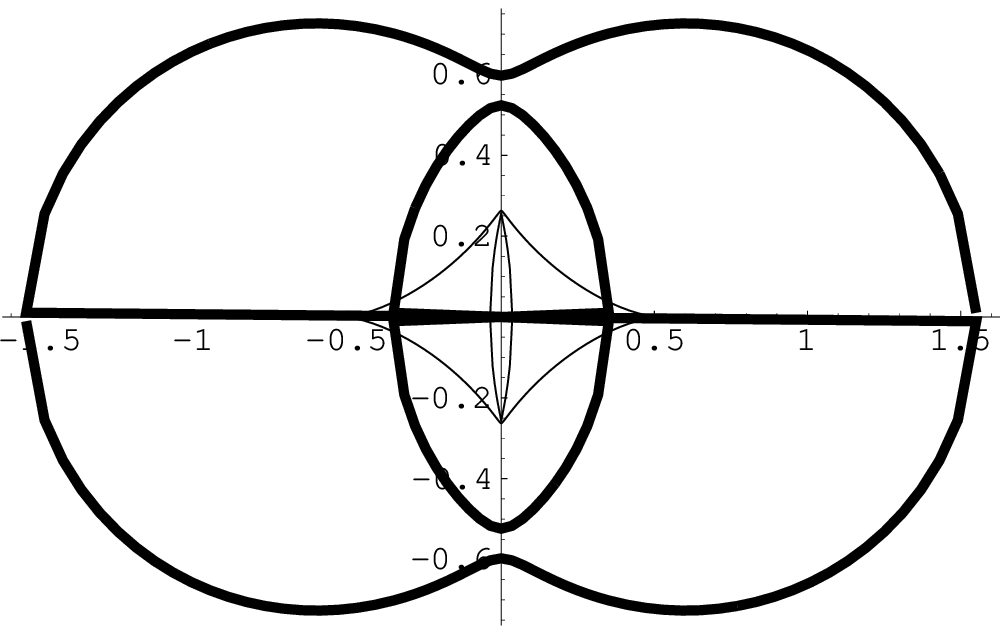}
\special{hscale=80 vscale=80 voffset=-40 hoffset=40 psfile=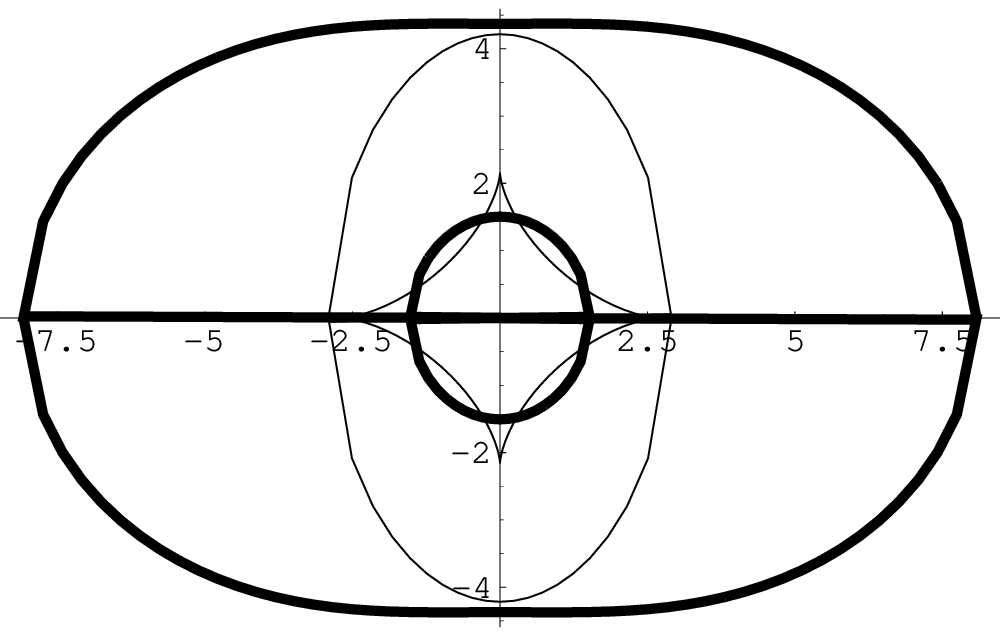}
\caption{Shape of the caustic lines (thick lines) and critical lines
(thin lines) for an elliptic potential and for different values of the
central potential. 
}
\label{EllipPot}
\end{figure}

The simplest assumption beyond the spherically symmetric models
is to introduce an ellipticity $\epsilon$ in the mass distribution
(Kassiola \& Kovner 1993),
\beq{
\varphi={\varphi_0 \sqrt{1+r_{\rm em}^2/r_c^2}}\ \ 
{\rm with}\ \ r_{\rm em}^2={x^2\over (1-\epsilon)^2}+{
y^2\over (1+\epsilon)^2}.\label{PotEll}
}
To understand the physics it induces one should introduce the caustics
and critical lines. The {\sl critical lines} are the location on the image
plane of the points of infinite magnification. 
The {\sl caustics} are the location of these points on the source plane. 
These points are determined by the lines on which $\det(\mA^{-1})=0$.
It means that arcs are along the critical lines and that they are
produced by background galaxies 
that happen to be located on the caustics. On Fig. \ref{EllipPot}
one can see the shapes of the critical lines and caustics for
different depth of the potential (\ref{PotEll})
(or equivalently for different positions of the source plane).

Eventually the reconstruction of galaxy cluster mass maps
requires the use of more complicated models and it can be necessary 
to perform non-parametric mass reconstructions. Recent
results have been obtained by AbdelSalam et al. (1997) for few clusters.

\section{The weak lensing regime}

In this section I consider the possibility of using the lens effects
to probe the large-scale structures of the Universe. The difficulty
is here that the distortion induced by the lenses can be very small.
The projected potential should then be reconstructed with a
statistical analysis 
on the deformation measured on a lot of background objects. Let me
define more precisely the different regimes.

\subsection{The mathematical description of the weak lensing regime}

Depending on the magnitude of the lens effect, different regimes are
possible:
\begin{itemize}
\item The strong regime is such that several light path are possible
between the sources and the observer. It thus induces multiple images,
and the images of background galaxies are often extremely
distorted: this is the regime of the giant arcs. The one that has
been investigated so far.
\item The regime of arclets correspond to a case where 
there are no multiple images although  a significant
distortion of the background objects can be observed.
\item The weak lensing regime corresponds to cases of modest distortion
(typically a few \%). Such effect cannot be detected with a single object
and therefore should be measured in a statistical way, by
averages over a large number of background galaxies.
\end{itemize}


In all cases the displacement field is not directly observable.
In the weak lensing regime, the deformation only is measurable. 
For slightly extended objects such as background galaxies
the deformation in shape is induced by the variations of the
displacement field. It can actually be described by the 
amplification matrix, the components of its inverse are in general written,
\beq{
\mA^{-1}=\left(
\begin{array}{cc}
1-\kappa-\gamma_1&-\gamma_2\\
-\gamma_2&1-\kappa+\gamma_1
\end{array}
\right),
}
taking advantage of the fact that it is a symmetric matrix.
The components of this matrix are expressed in terms of
the convergence, $\kappa$, (a scalar
field) and the shear, $\gamma$ (a pseudo vector field) with
\beq{
\kappa={1\over 2} \nabla^2 \psi; 
\ \ \  \gamma_1={1\over 2}(\psi_{,11}-\psi_{,22}) \ ; 
\ \ \ \gamma_2=\psi_{,12}, \ \ {\rm with}\ \ 
\psi=2{D_{LS}\over D_{OS}\,D_{OL}}\phi.
}
The convergence describes the linear change of size 
and the shear describes the deformation.
The consequences of such a transform can be decomposed in two aspects:
\begin{itemize}
\item The magnification effect.
Lenses induce a change of size of the objects. As the surface brightness
is not changed by this effect, the change of surface induces a
direct magnification effect, $\mu$. This magnification
is directly related to the determinant of $\mA$ so that,
\beq{
\mu=\det(\mA)=1/\left[(1-\kappa)^2-\gamma^2\right].
}
\item The distortion effect.
Lenses also induce a change of shape of the background objects. 
The eigenvalues of the
matrix $\mA^{-1}$ determine the direction and amplitude of such a 
deformation. 
\end{itemize}

\subsection{The magnification effect}

In the weak lensing regime,
the observational consequences of the magnification effect is a 
combination of a change the apparent area of the objects, 
that makes their detection
easier, and their dilution (the total area is enlarged as well). 
The mean local number density of galaxies is then  related to the
slope of the galaxy counts, $\alpha$, through
\beq{
n(\gamma)=n_g\,\mu^{2.5\alpha-1}\ \ 
{\rm with}\ \ \alpha={\d\log N\over\d m},
}
where $m$ is the apparent magnitude (in a given band)
and $n_g$ is the mean number density of galaxies in the
absence of lenses. Whether the number density of galaxies
increases or decreases in magnified area thus depends on $\alpha$
and consequently on the selected population of objects.

Such an effect has been advocated (Broadhurst 1995 and Broadhurst et
al. 1995) as a way to detect the lens effect. In general this 
method suffers from the fact that
the background galaxies have intrinsic number density fluctuations.
It is therefore more appropriate for mapping the
mass fluctuations in galaxy clusters where
the magnification effect is large enough to dominate. 
In galaxy clusters it is 
a cheap way to map the mass profiles. More sophisticated analysis
allow even to have access to the cosmological constant 
$\Lambda$ by probing the extension of the depletion area (Fort et
al. 1997, see contribution of Y. Mellier).

\subsection{The galaxy shape matrices to measure the distortion field}

The distortion effects change the shape of the background objects. 
The objects
appear elongated along the eigenvalues of the amplification matrix.
When background objects are only moderately extended
(this excludes the case of arcs), 
their shapes can be described by the matrix,
\beq{
\mS\equiv\disp{\int\d^2\theta\,\theta_i\,\theta_j\,\mI(\theta)}.
}
It is easy to relate the shape matrix in the source plane to the one
in the image plane. This is obtained by a simple change of variable
that uses the fact that the surface brightness of the objects
is not changed. It implies
\beq{
\mS^S={\mA^{-1}\cdot\mS^I\cdot\mA^{-1}}.
}
By averaging over the shape matrices in the source plane, assuming
the intrinsic shape fluctuations are not correlated, one can
eventually get\footnote{See Mellier, these proceedings, for a more
detailed presentation of the data analysis techniques.} 
the value of $\mA^{-1}/\sqrt{\det\mA^{-1}}$. The combination we have
access to is
totally independent of the amplification factor. As a consequence, the
quantity which is measurable is the reduced shear field,
\beq{
\vg={\vec{\gamma}/(1-\kappa)}.
}
This quantity identifies with $\gamma$ only in the limit of very weak
lensing (i.e. when $\kappa\ll1$).

\subsection{The construction of the projected mass density}

The elaboration of methods for reconstructing mass maps from
distortion fields is not a trivial issue. In a pioneering paper, Kaiser \&
Squires (1993) showed that this is indeed possible, at least
in the weak lensing regime. It is indeed not too difficult to show that (in the single lens approximation),
\beq{
\grad \kappa=-
\left(\begin{array}{cc}
\partial_1&\partial_2\\
-\partial_2&\partial_1
\end{array}\right)
\cdot
\left(\begin{array}{c}
\gamma_1\\
\gamma_2
\end{array}\right)
}
when $\kappa\ll1 $ and $\gamma_i\ll 1$. By simple Fourier transforms
it is then possible to recover $\kappa$ from a distortion map.

Such a method was further extended in many ways. Bartelmann, Schneider
and his  collaborators (Bartelmann et al 1996, Schneider 1995, Seitz
\& Schneider 1995, Seitz \& Schneider 1996, Seitz et al. 1998) have
worked in detail on the edge effects, the possibility of having
adaptive smoothing procedures, the use of maximum entropy method
etc... This is particularly important when structures of different
sizes and contrasts  are present at the same time.  Finally Kaiser
(Kaiser 1995) exhibited the relation between the local convergence and
the distortion field $\vg$ which is valid in all regimes. This
relation reads 
\beq{ \grad \log(1-\kappa)= \left(\begin{array}{cc}
1-g_1&g_2\\ g_2&1+g_1
\end{array}\right)^{-1}
\cdot
\left(\begin{array}{cc}
\partial_1&\partial_2\\
-\partial_2&\partial_1
\end{array}\right)
\cdot
\left(\begin{array}{c}
g_1\\
g_2
\end{array}\right).
}
This is a non-linear and non-local relationship. 

The first reconstruction of a mass map of a galaxy cluster has been
done on MS1224 by Fahlman et al. (1994). Many other
reconstructions have now been done or are under preparation
(see Mellier 1998 and these proceedings).

\section{The weak lensing as a probe of the Large-Scale Structures}

\subsection{The large-scale structures}

The idea of probing the large-scale structures with 
gravitational effects is very attractive. The gravitational survey offers
indeed a unique way of probing the mass concentrations in the
Universe since, contrary to galaxy survey, it can provide us with 
mass maps of the Universe that are free of any bias. 
Its interpretations in terms of cosmological parameters would 
then be straightforward and independent on hypothesis on galaxy or cluster
formation schemes. 

The physical mechanisms are the same in the context of large-scale
structures and the source for the gravitational effects is the
gravitational potential $\phi$ given by Eq. (\ref{PoissonCosmo}).
It is important to remember that the source term of this
equation is $\rhob(t)\,\deltam(t,\vx)$.
The density contrast $\deltam(\vx)$ is usually written in terms
of its Fourier transforms,
\beq{
\deltam(\vx)=\int{\d^3\vk\over
(2\pi)^{3/2}}\,\deltam(\vk)\,D_+(t)\,\exp(\ii\vk.\vx).
}
The density field is then entirely defined by the statistical 
properties\footnote{For a detailed introduction to large-scale
structure formation theory and phenomenology  
see lecture notes of Bertschinger, 1996, and
these proceedings.}
 of the random variables $\deltam(\vk)$. At large enough
scale the field is (almost) Gaussian (at least for Gaussian initial 
conditions which is the case in inflationary scenarios).
The amplitude of the fluctuations grows with time in linear theory
in a known way $D_+(t)$. This function is simply 
proportional to the expansion factor for an Einstein-de Sitter
Universe.
 
The variables are then entirely
determined by the power spectrum $P(k)$,
\beq{
\mg\deltam(\vk)\deltam(\vk')\md=\delta_{\rm Dirac}(\vk+\vk')\,P(k).
}
The cosmological model is therefore completely 
determined by the power spectrum,
$\Omega$ and $\Lambda$ as long as 
the the dark matter distribution is concerned.

\begin{figure*}
\centerline{
\psfig{figure=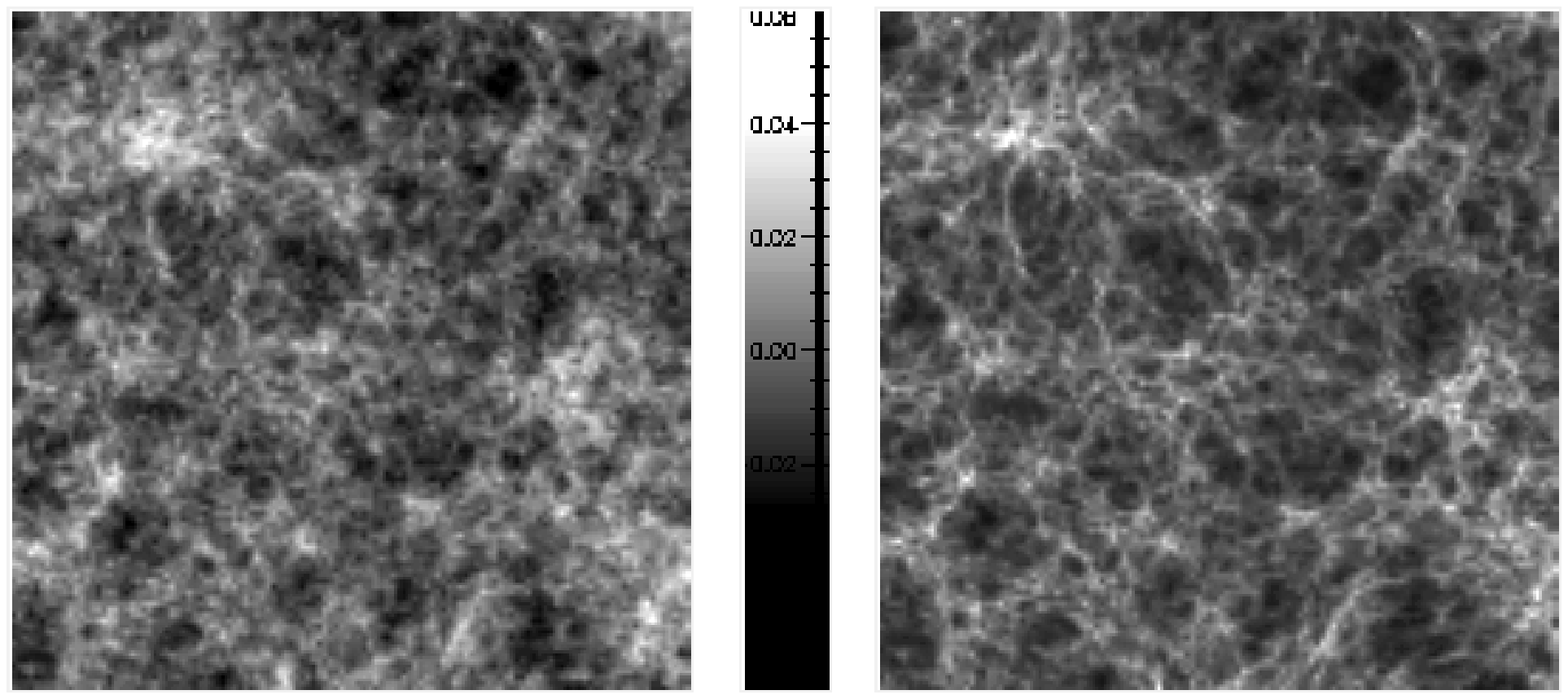,height=5.5cm}}
\centerline{
\psfig{figure=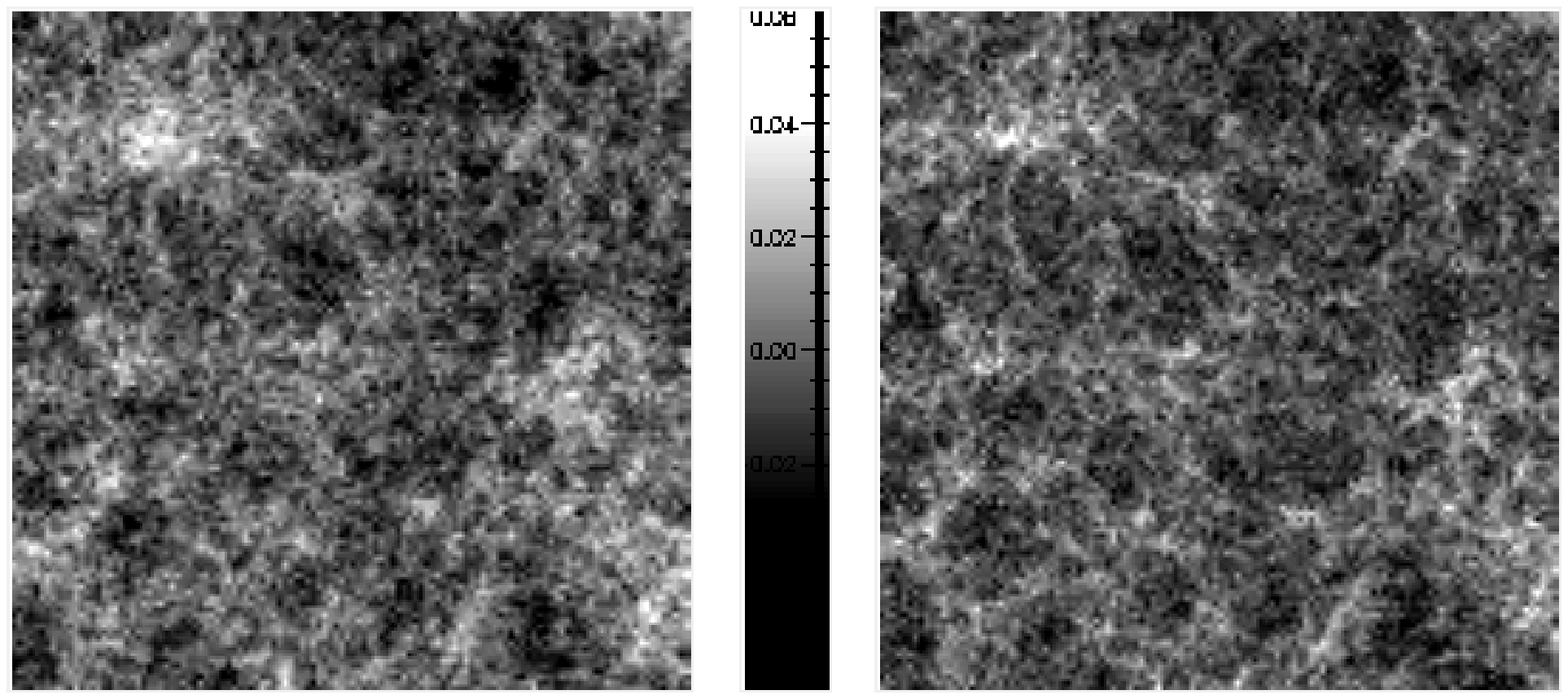,height=5.5cm}}
\caption{Example of reconstructions of
projected mass maps. The top panels show the initial noise-free
$\kappa$ map for either $\Omega=1$ (left panel) or $\Omega=0.3$ (right panel)
with the same underlying linear random field and
the same rms distortion.
The bottom panels show the reconstructed $\kappa$ maps with noise included in
the shear maps.
The maps cover a total area of  25 degrees$^2$. Each pixel has an
angular size of
2.5 arcmin$^2$ and averages the shear signal expected from deep CCD exposures
(about 30 galaxy/arcmin$^2$). The sources are assumed 
to be all at redshift unity and to have a realistic intrinsic
ellipticity distribution. Such a survey is easily accessible to 
MEGACAM at CFHT.
The precision with which the images can be reconstructed and the
striking differences between the two cosmological models
demonstrate the great interest such a survey would have.
}
\label{kappa_maps}
\end{figure*}

\subsection{The relation between the local convergence and the local
density contrast} 

The relation between the convergence and the local density contrasts
in the local universe can be derived easily from
Eqs. (\ref{PoissonCosmo}, \ref{AmpSachs}),
\beq{
\kappa(\gamma)={3\over2}\Omega_0
\int\d z_s\,n(z_s)\int\d\chi\,{\mD(\chi_s,\chi)\,\mD(\chi)\over \mD(\chi_s)}
\,\delta_{\rm mass}(\chi,\gamma)\,(1+z)\label{kappa}.
} 
In this relation the redshift distribution of the sources
in normalized so that,
\beq{
\int\d z_s\,n(z_s)=1.
}
All the distances are expressed in units of $c/H_0$.
The relation (\ref{kappa}) is then totally dimensionless.

\subsection{The efficiency function}

\begin{figure}
\vspace{6 cm}
\special{hscale=70 vscale=70 voffset=-40 hoffset=40 psfile=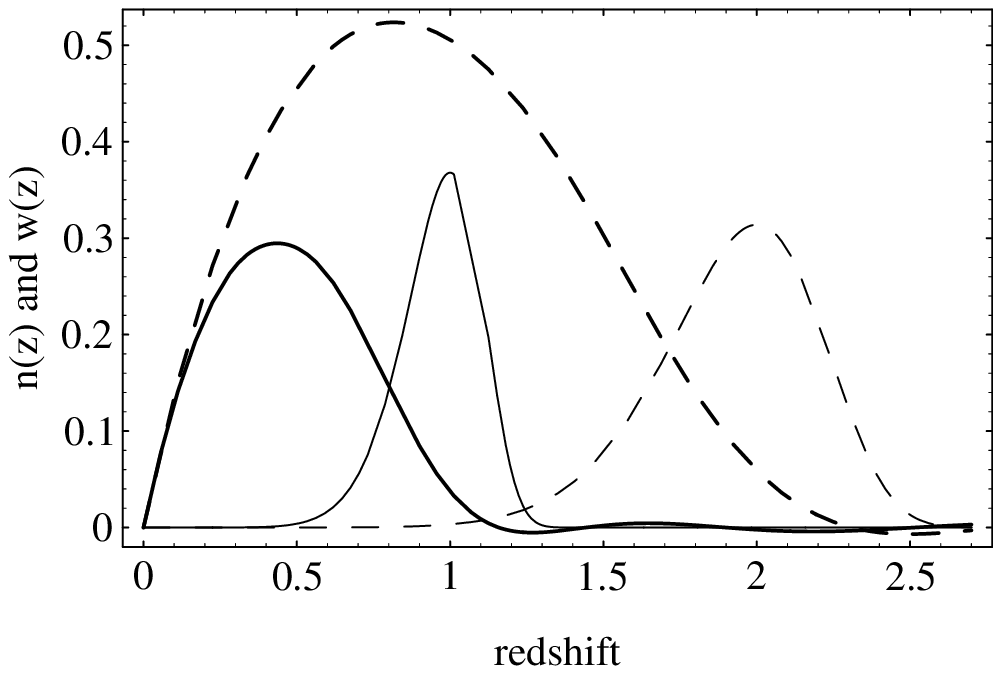}
\caption{shape of the efficiency function, $w(z)$ (thick lines),
for two different hypothesis on the shape of
the redshift distribution of the sources (thin lines).
}
\label{effdez}
\end{figure}

It is convenient to define the efficiency function, $w(z)$,
with
\beq{
w(z)={3\over 2}\Omega_0\int\d z_s\,n(z_s)\,
{\mD(\chi_s,\chi)\,\mD(\chi)\over \mD(\chi_s)}
\,(1+z)\label{eff}
}
On Fig. \ref{effdez}
one can see the shape of the efficiency function for different
hypothesis for the source distribution. Obviously the further the
sources are the more numerous the lenses that can be detected are, and
the larger the effect is.

\subsection{The amplitude of the convergence fluctuations}

From this equation
it is obvious that the amplitude of $\kappa$ is directly proportional of the
density fluctuation amplitude and that the two point correlation function
of the $\kappa$ field is related to the shape of the density
power spectrum. In the following its amplitude is
parameterized with $\sigma_8$ which is the r.m.s. of the density
contrast in a sphere of radius $8\,h^{-1}$Mpc.
The relation (\ref{kappa}) 
also shows that $\kappa$ depends on the
cosmological parameters. There is a significant dependence in the 
expression of the distances but the dominant contribution comes from
the overall  $\Omega_0$ factor.

The amplitude of the fluctuations of $\kappa$ depends on the angular scale 
at which the convergence map is filtered.
We can introduce the filtered convergence $\kappa_{\theta}$, with
\beq{
\kappa_{\theta}(\gamma)=
\int\d^2\gamma'\,\kappa(\gamma+\gamma')\,W_{\theta}(\gamma').
}
It is convenient to introduce the Fourier transform of the
window function $W(k)$. This function is,
\beq{
W(k)=2{J_1(k)\over k},
}
where $J_1$ is the Bessel function, in case of a 
angular top-hat filter.
Then the filtered convergence reads,
\bea{
\kappa_{\theta}(\gamma)
=&\disp{\int\d \chi\,
w(z)\int{\d^2\vk_{\perp}\over 2\pi}\,{\d k_r\over (2\pi)^{1/2}}\,
\delta(\vk)\,D_+(z)\,}\times\nonumber\\
&\disp{\exp\left[\ii\vk_r\chi(z)+\ii\vk_{\perp}.\gamma\mD(z0)\right]\,
W\left[k_{\perp}\theta\mD(z)\right]},
}
where the wave vector $\vk$ has been decomposed in two parts
$k_r$ and $\vk_{\perp}$ that are respectively along the line of sight
and perpendicular. The computation of the r.m.s. of $\kappa_{\theta}$
is analytic in the small angle approximation only. In such an
approximation we have, 
\beq{
\theta\ll 1\ \ \Rightarrow \ \ k_r\ll k_{\perp}\ \ \Rightarrow \ \ 
P(k)\,D_+^2(z)\approx P(k_{\perp}).
}
Eventually the variance reads,
\beq{
\mg\kappa^2_{\theta}\md=\int\d \chi\,w^2(\chi)\,\int
{\d^2\vk\over 2\pi}\,P(k)\,W^2(k_{\perp}\theta\mD).
}
For realistic models of the power spectrum (e.g. Baugh and
Gazta\~naga, 1995), the numerical result is (Bernardeau et al. 1997),
\beq{
\mg\kappa^2_{\theta}\md^{1/2}\approx 0.01\ \sigma_8\ \Omega_0^{0.8}\ 
z_s^{0.75}\ \left({\theta\over 1\deg}\right)^{-(n+2)/2}.\label{w2res}
}
To be noticed is the dependence on the redshift of the sources.
This was noticed by Villumsen (1996) who pointed out that the
$\Omega_0$ dependence is roughly given by the $\Omega$
value at the redshift of the sources. These results are slightly affected by 
the introduction of the non-linear effects
in the shape of the power spectrum (Miralda-Escud\'e 1991,
Jain \& Seljak 1997).

\subsection{The expected signal to noise ratio}

Are the effects from large-scale structures measurable?
It depends on the number density of background objects for which
the shape matrices are measurable. In current deep galaxy survey 
the typical mean number density of objects is about 50 arcmin$^{-2}$. 
The precision of the measured distortion at the
degree scale is then about,
\beq{
\Delta_{\rm noise}\kappa={0.3\over \sqrt{50\ 60^2}} \approx 10^{-3},
}
for an intrinsic ellipticity of sources of about $0.3$.
This number is to be compared with the
expected amplitude of the signal coming from the large-scale
structures,
about 1\% according to Eq. (\ref{w2res}) (see also earlier computations by
Blandford et al. 1991, Miralda-Escud\'e 1991,
Kaiser 1992).
This makes such detection a priori possible with a signal to noise
ratio around 10 (provided the instrumental
noise can be controlled down to such a low level).

\subsection{Separate measurements of $\Omega$ and $\sigma_8$}

In Eq. (\ref{w2res}) one can see that the amplitude of the
fluctuations
depend both on $\sigma_8$ and on $\Omega_0$.
A question that then arises is whether it is possible to
 separate the amplitude of the power spectrum from the cosmological
parameters. A simple examination of the equation (\ref{kappa})
shows that it should be the case,
because, for a given value of $\sigma_{\kappa}$, the 
density field is more strongly evolved into the non-linear 
regime when $\Omega_0$ is low.

\begin{figure}
\vspace{14 cm}
\special{hscale=60 vscale=60 voffset=240 hoffset=-20 psfile=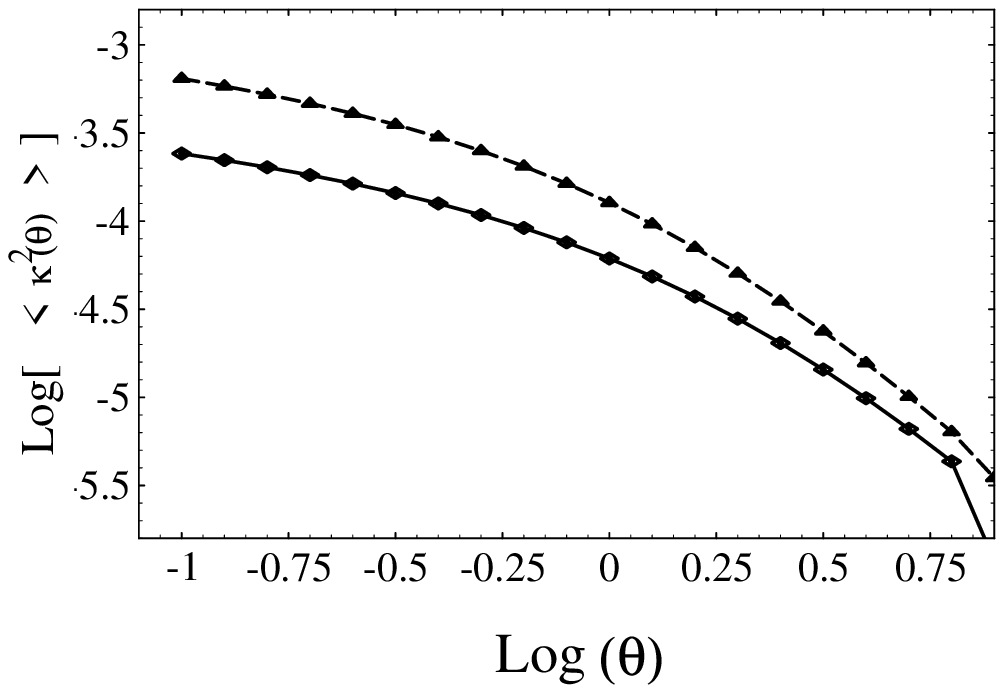}
\special{hscale=60 vscale=60 voffset=120 hoffset=-10 psfile=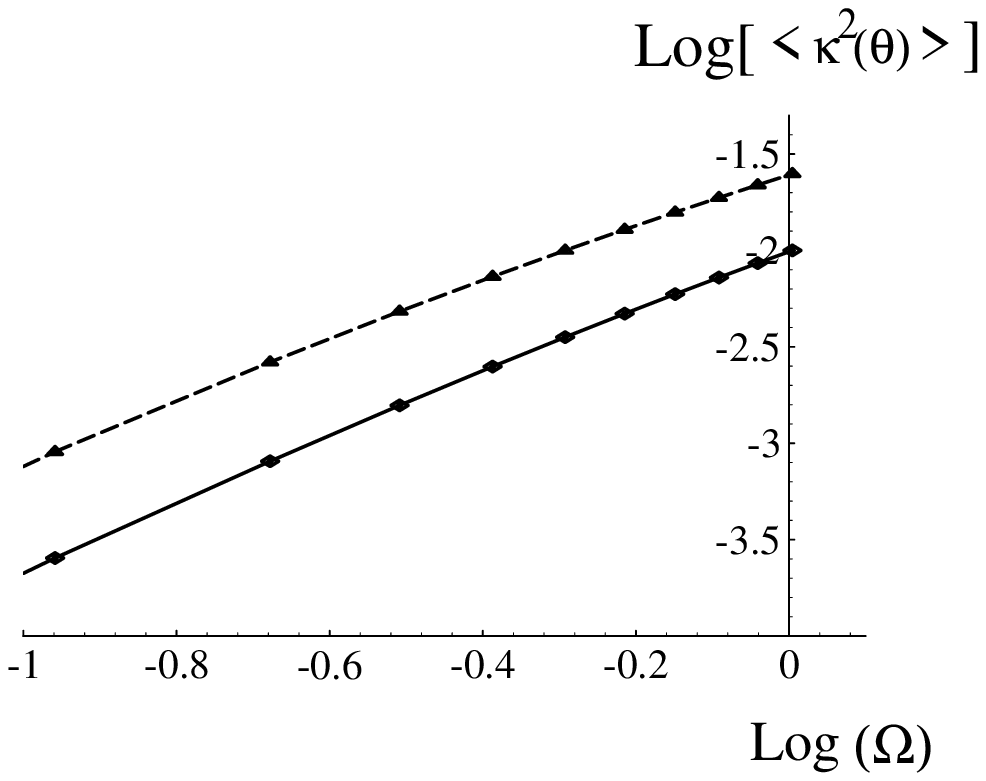}
\special{hscale=60 vscale=60 voffset=-20 hoffset=-20 psfile=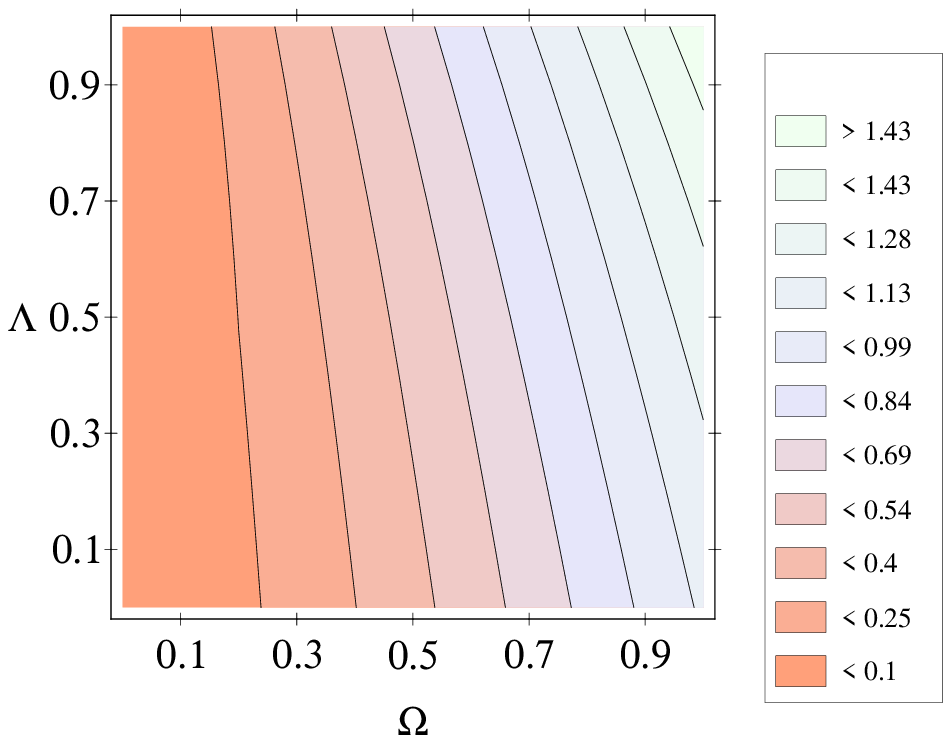}
\special{hscale=60 vscale=60 voffset=240 hoffset=170 psfile=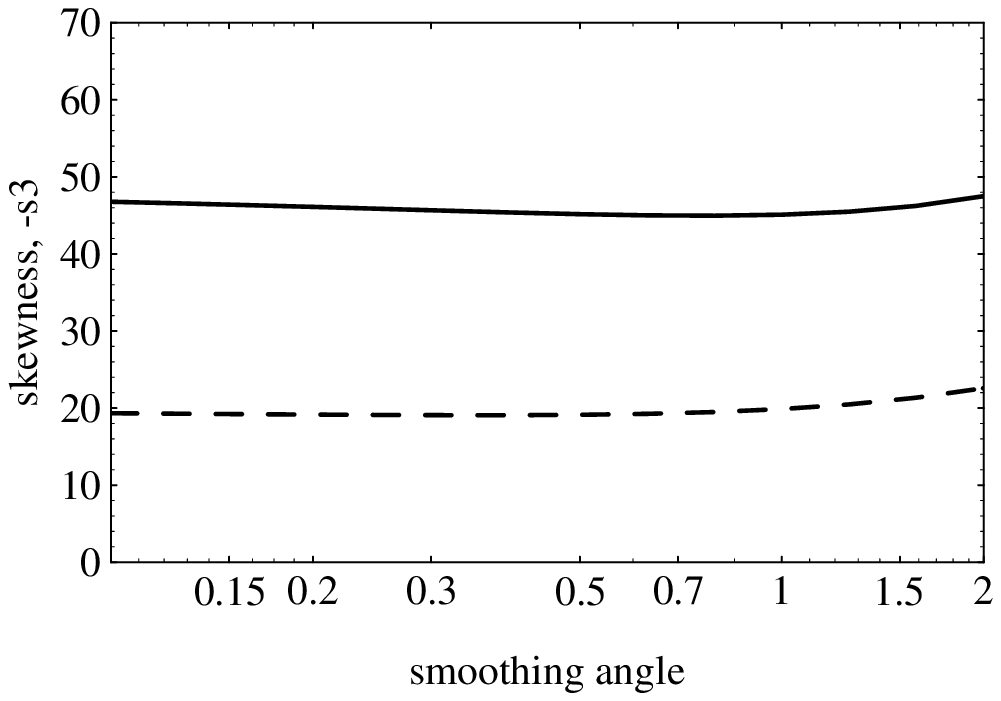}
\special{hscale=60 vscale=60 voffset=120 hoffset=190 psfile=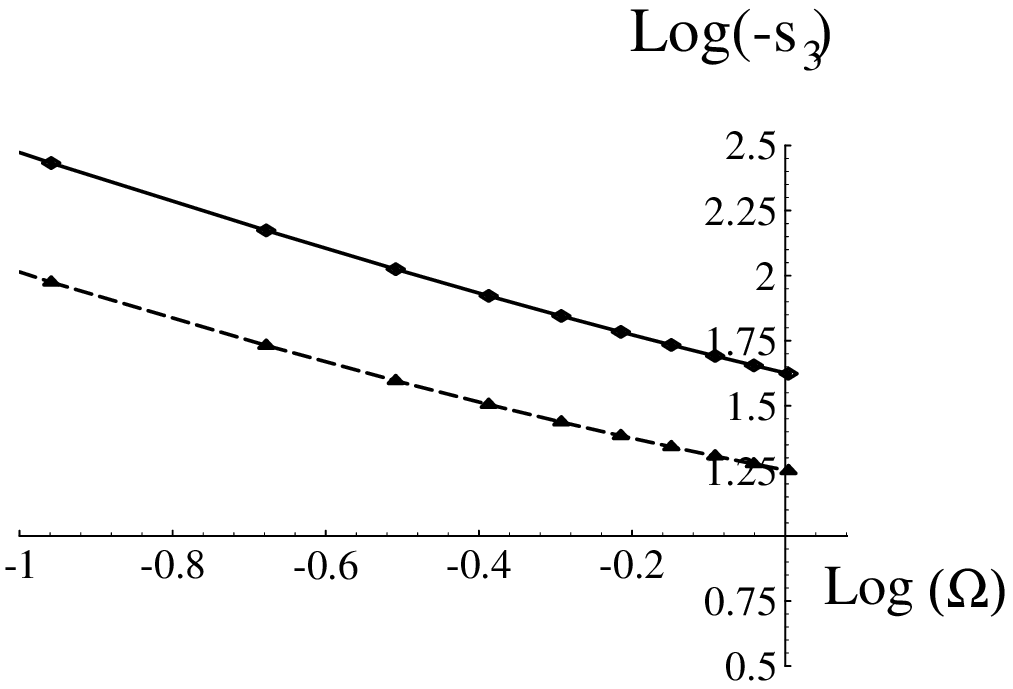}
\special{hscale=60 vscale=60 voffset=-20 hoffset=170 psfile=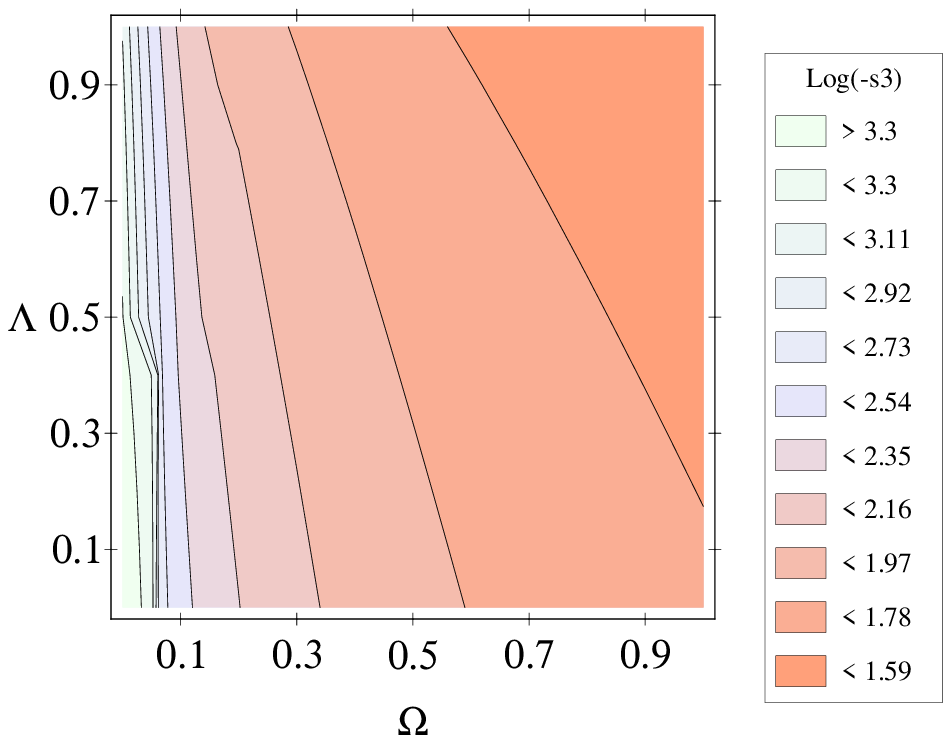}
\caption{Perturbation Theory results (Bernardeau et al. 1997)
 obtained for the width (left panels) and skewness
(right panels) of the probability distribution function of the
local convergence. The results are plotted as a function of the
angular scale in the top panels, of $\Omega_0$ in the middle
ones. The solid lines correspond to source redshifts of 1, and
the dashed lines to redshifts of 2.
In the bottom panels the iso-values of the width and the
skewness are plotted in the $\Omega_0$-$\Lambda$ plane (for
source redshifts of 1). It shows that the dependence on
$\Omega_0$ is slightly degenerate with $\Lambda$. 
}
\label{kappaStats}
\end{figure}

The consequences of this are two fold. The nonlinearities
change the angular scale at which the non-linear dynamics starts 
to amplify the growth of structures. 
This effect was more particularly investigated
by Jain \& Seljak (1997) who showed that the emergence of
the nonlinear regime 
is apparent in the shape of the angular two-point function. 
This effect is however quite subtle since 
it might reveal difficult to separate
from peculiarities in the shape of the initial power spectrum.

The other aspect is that nonlinear effects induce 
non-Gaussian features due to mode couplings. 
These effects have been studied extensively in Perturbation
Theory. Technically one can write the local density contrast
as an expansion with respect to the initial density fluctuations,
\beq{
\deltam(\vx)=\deltam^{(1)}(\vx)+\deltam^{(2)}(\vx)+\dots
}
where $\deltam^{(1)}(\vx)$ is proportional to the
initial density field (this is the term we have considered so far),
$\deltam^{(2)}(\vx)$ is quadratic, etc.
Second order perturbation theory provides us with the
expression of $\deltam^{(2)}(\vx)$ (there are many references
for the perturbation theory calculations, Peebles 1980,
Fry 1984, Goroff et al. 1986,  
Bouchet et al. 1993 for the $\Omega$ dependence of this result),
\bea{
\deltam^{(2)}(t,\vx)=&\disp{\int{\d^3\vk_1\over (2\pi)^{3/2}}\,
{\d^3\vk_2\over (2\pi)^{3/2}}\,D_+^2(t)\,\deltal(\vk_1)\deltal(\vk_2)\,
\exp[\ii(\vk_1+\vk_2)\cdot\vx]}\times\nonumber\\
&\disp{\left[{5\over 7}+{\vk_1\cdot\vk_2\over k_1^2}+{2\over 7}
{(\vk_1\cdot\vk_2)^2\over k_1^2\,k_2^2}\right]},\label{delta2}
}
where $\deltal(\vk)$ 
are the Fourier components of the {\em linear} density field.
It behaves essentially as the square of the linear term, with 
a non-trivial geometric function that contains the non-local effects
of gravity.

Equivalently it is possible to expand the local convergence in terms
of the initial density field,
\beq{
\kappa(\gamma)=\kappa^{(1)}(\gamma)+\kappa^{(2)}(\gamma)+\dots
}
The apparition of a non-zero $\kappa^{(2)}$ induces non-Gaussian
effects that can be revealed for instance by the computation
of the skewness, third moment, of $\kappa_{\theta}$
(Bernardeau et al. 1997),
\beq{
\mg\kappa_{\theta}^3\md=\mg\left(\kappa_{\theta}^{(1)}\right)^3\md+
3\,\mg\left(\kappa_{\theta}^{(1)}\right)^2\kappa_{\theta}^{(2)}\md+\dots
}
The actual dominant term of this expansion is 
$3\,\mg\left(\kappa^{(1)}\right)^2\kappa^{(2)}\md$ since the first
term vanishes for Gaussian initial conditions.
For the computation of such term one should
plug in Eq. (\ref{kappa}) the expression of $\deltam^{(2)}$
in Eq. (\ref{delta2}) and do the computations in the small
angle approximation (and using specific properties of the
angular top-hat window function, Bernardeau 1995).

Eventually perturbation theory gives the following result
for a realistic power spectrum (Bernardeau et al. 1997),
\beq{
s_3(\theta)\equiv{\mg\kappa^3_{\theta}\md\over
\mg\kappa^2_{\theta}\md^2}=
40\ \Omega_0^{-0.8}\ z_s^{-1.35}.\label{s3}
}
The origin of this skewness is relatively simple
to understand: as the density field enters the non-linear
regime the large mass concentrations tend to acquire a large 
density contrast in a small volume. This induces
rare occurrences of large negative convergences. The under-dense
regions tend on the other hand to occupy a large
fraction of the volume, but can induce only moderate
positive convergences. This mechanism is clearly visible on the maps
of figure (\ref{kappa_maps}). When the mean source redshift
grows the skewness diminishes since the addition of independent 
layers of large-scale structures tend to dilute the non-Gaussianity.

\begin{figure}
\centerline{
\psfig{figure=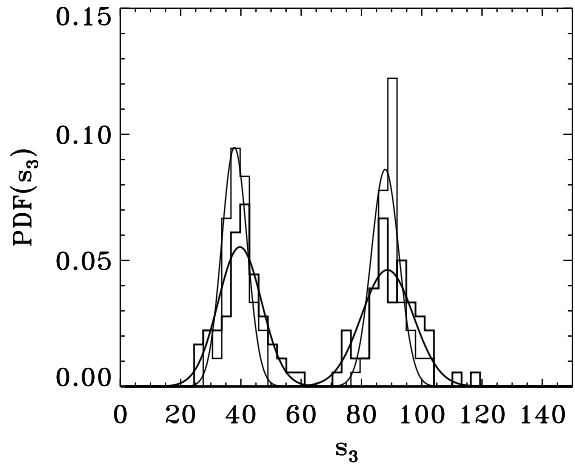,height=7cm}}
\caption{\label{Histo.s3}
Histograms of the values of $s_3$, top-hat filter, 
for $\Omega=1$ and $\Omega=0.3$
for a $5\times5$ degree survey (thick lines) and a $10\times10$ degree survey
(thin lines).
}
\end{figure}

What the Eq. (\ref{s3}) demonstrates is that distortion maps
can be used to determine the cosmic density parameter, $\Omega_0$,
provided the redshift distribution of the sources is well
known. The hierarchy exhibited in this relation is also a direct
consequence of the hypothesis of Gaussian initial conditions. 
Such a hierarchy has been observed for instance in galaxy catalogues
(see  Bouchet et al. 1993 for results in the
IRAS galaxy survey). It can be very effective in excluding
models with non-Gaussian initial conditions (see the
attempt of Gazta\~naga \& M\"ah\"onen 1996).
To be more precise I present the actual histograms of the measured
skewness in numerical simulations (Fig. \ref{Histo.s3}) 
which clearly demonstrate that the two
cosmologies are easily separated.
One can see that the scatter in $s_3$ is roughly the same in the two cases
and that the difference in the relative precision is due to the
differences in the expectation values. 

\begin{figure}
\centerline{
\psfig{figure=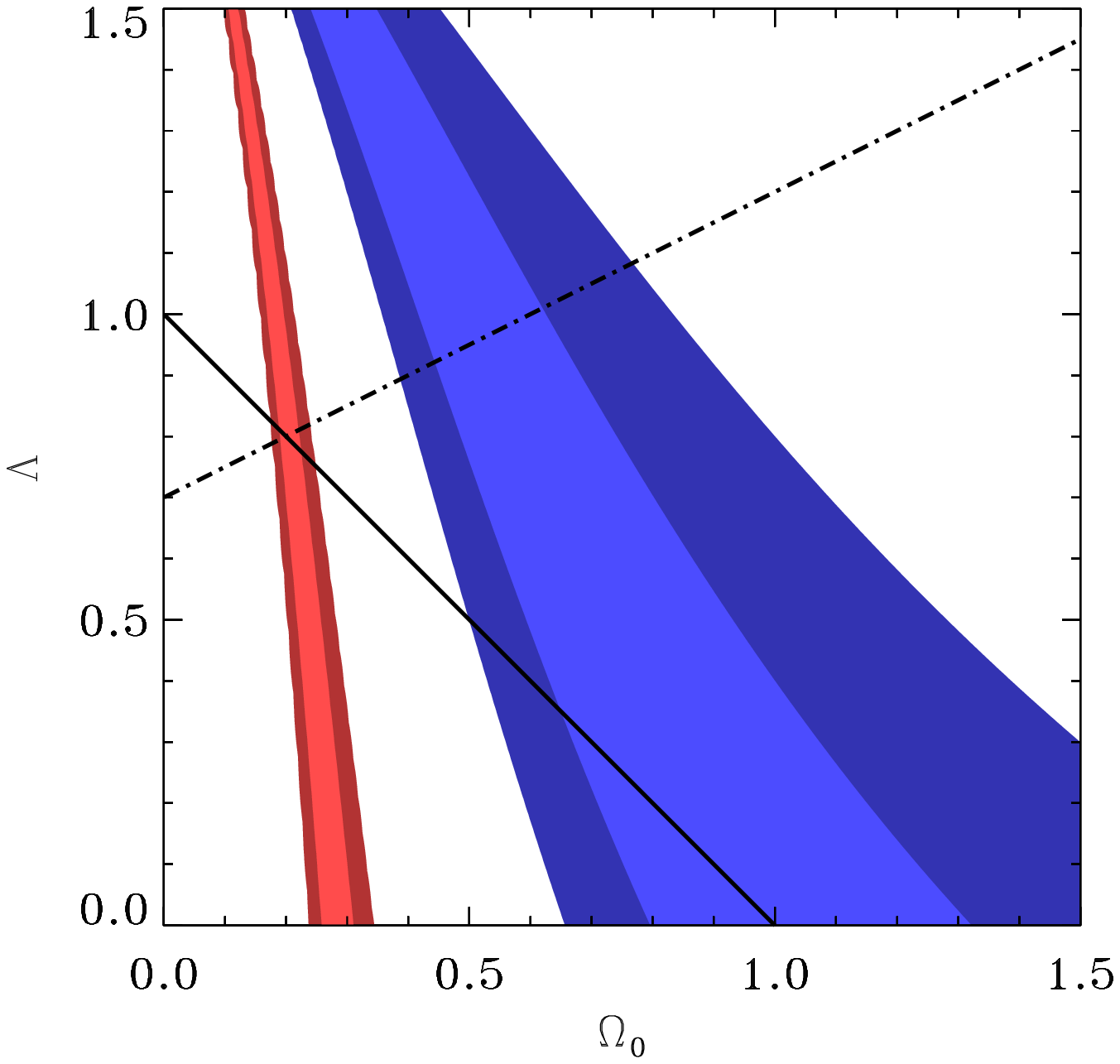,width=7cm}}
\centerline{
\psfig{figure=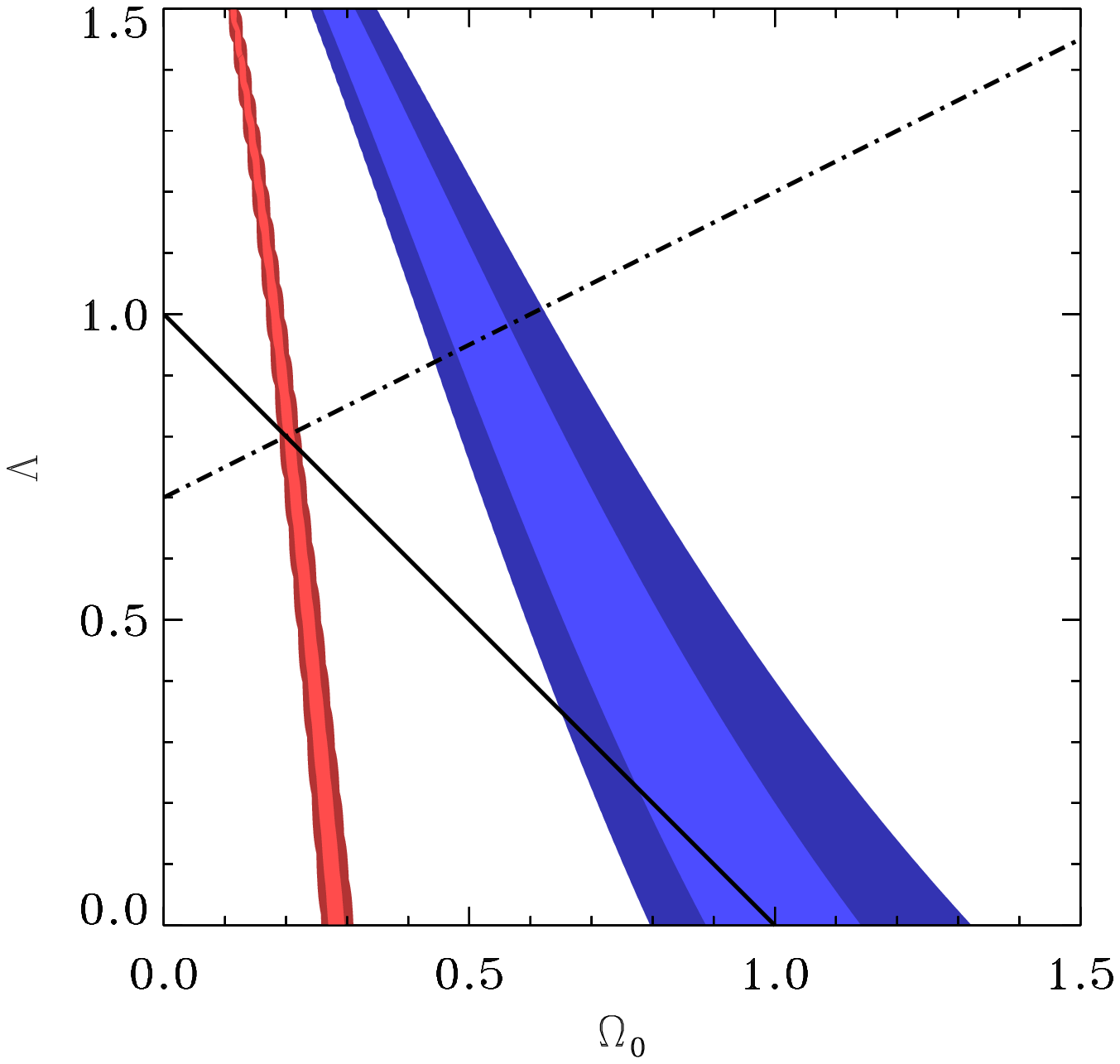,width=7cm}}
\caption{\label{OmLambda.ps} Constraints that can be brought 
by weak lensing survey in an $\Omega_0-\Lambda$ plane. 
The grey bands are the location of the 1 and 2-$\sigma$
locations (respectively darker and lighter bands)
allowed by a measured skewness that would be obtained
with either $\Omega_0=0.3$ (left bands) or $\Omega_0=1$ (right bands).
The solid straight lines corresponds to a zero curvature universe, and
the dot-dashed lines to a fixed acceleration parameter, $q_0$. 
The panels correspond to survey
of either $5\times5$ (top) or $10\times10$ degrees (bottom).
}
\end{figure}

The validity of Eq. (\ref{s3}) has been confirmed numerically by
Gazta\~naga \& Bernardeau (1998), who showed it is
valid for scales above a few tens of arcmins.
A non-zero skewness has also been observed in the numerical experiment
of Jain et al. (private communication, in preparation) and
van Waerbeke et al. (1998).
Large angular convergence maps can therefore provide new means
for constraining fundamental cosmological parameters.
Numerical results show that in maps of 
$5\times 5$ square degrees it is reasonable to expect a precision
of a few percent on the normalization and 
about $5\%$ to $10\%$ on the cosmological density parameter depending
on the underlying cosmological scenario (see Fig. \ref{OmLambda.ps}).

\subsection{Prospects}

From an observational point of view,
the investigation of the large-scale structures of the Universe with
gravitational lenses is in a very preliminary stage. After an early
claim by Villumsen (1995), a direct evidence
of the detection distortion signal of gravitational
origin has been reported recently by Schneider et al (1997).

There are at present many studies, either theoretical or numerical,
that aim to examine all possible systematic errors (Bonnet \& Mellier
1995, Kaiser et al. 1995), to optimize the data analysis concepts
(such as the pixel autocorrelation function by van Waerbeke et
al. 1997) and the scientific interpretations of the resulting mass
maps (Bernardeau 1998, Bernardeau et al. 1997,  Seljak 1997, van
Waerbeke et al. 1998).  A few observational surveys are now emerging,
\begin{itemize}
\item the ESO key program jointly done by MPA and IAP;
\item the DESCART project, part of the scientific program
of the wide field CCD camera to be installed at the CFHT.
\end{itemize}

\acknowledgements
The author thanks Y. Mellier for innumerable discussions on the
lens physics and IAP for its hospitality.

\end{document}